\documentclass[12pt]{iopart}

\usepackage{graphicx}
\usepackage{amstext,amssymb}
\usepackage{color}
\usepackage{booktabs}
\usepackage[normalem]{ulem}
\usepackage{cancel}

\def\lsim{\raise0.3ex\hbox{$\;<$\kern-0.75em\raise-1.1ex
\hbox{$\sim\;$}}}
\def\gsim{\raise0.3ex\hbox{$\;>$\kern-0.75em\raise-1.1ex
\hbox{$\sim\;$}}}

\begin{document}

\title[NSI]
{Non standard neutrino interactions: current status and future prospects}

\author{O. G. Miranda}

\address{Departamento de F\'{\i}sica, Centro de Investigaci{\'o}n y de 
  Estudios Avanzados del IPN, Apdo. Postal 14-740 07000 
  M\'exico, D F, M\'exico}
\ead{omr@fis.cinvestav.mx}

\author{H. Nunokawa}

\address{Departamento de F\'{\i}sica, Pontif\'{\i}cia Universidade
  Cat\'olica do Rio de Janeiro, C. P. 38071, 22452-970, Rio de
  Janeiro, Brazil}
\ead{nunokawa@puc-rio.br}

\begin{abstract}
Neutrino oscillations have become well-known phenomenon;
the measurements of 
neutrino mixing angles and mass squared differences
are continuously improving.
Future oscillation experiments will 
eventually determine the remaining unknown neutrino parameters, 
namely, the mass ordering, normal or inverted, 
and the CP-violating phase.
On the other hand, the absolute mass scale of neutrinos could be
probed by cosmological observations, single beta decay as well as by
neutrinoless double beta decay experiments.  
Furthermore, the last one may shed light on the nature of neutrinos, 
Dirac or Majorana, by measuring the effective Majorana mass of neutrinos.
However, the neutrino mass generation mechanism remains unknown.  
A well-motivated phenomenological approach to search for new physics, in
the neutrino sector, is that of non-standard interactions.  
In this short review, the
current constraints in this picture, as well as the perspectives from
future experiments, are discussed.
\end{abstract}

\maketitle
\section{Introduction}

The last two decades have seen the great success of many neutrino
experiments, in particular, those that have contributed to the first
observations of the various types of neutrino
oscillation~\cite{Fukuda:1998mi,Ahmad:2002jz,
  Eguchi:2002dm,Abe:2011sj,Abe:2011fz,An:2012eh,Ahn:2012nd}.
The ``standard'' oscillation parameters, defined in the framework of
the three active neutrinos, have been determined with significant
accuracy~\cite{Agashe:2014kda} apart from the yet unknown mass
ordering, or the sign of the mass squared difference relevant for
oscillation of atmospheric neutrinos, as well as the CP-violating
phase, the most elusive parameter.
Non-oscillation experiments that have measured the neutrino cross
section with high accuracy, also provided valuable information for
neutrino physics~\cite{Fiorentini:2013ezn,Fields:2013zhk,Formaggio:2013kya}.
On the other hand, there are oscillation
experiments~\cite{Aguilar:2001ty,MiniBooNE,Kaether:2010ag,Abdurashitov:2009tn}
that have observed hints for neutrino oscillation into an additional
sterile neutrino state. 

So far, there is no experimental evidence that neutrinos
pose some non-standard properties beyond masses and mixing or some 
extra new interactions, different from the weak interaction, 
not described by the Standard Model (SM).
Such interactions, often called {\it non-standard interactions} 
(NSI) of neutrinos, if they exist, 
are interesting from a phenomenological
point of view, since they directly indicate the presence 
of some new physics beyond SM.

Possible presence of NSI was first pointed out by
Wolfenstein~\cite{Wolfenstein:1977ue,Wolfenstein:1979ni}, followed by
the works done in
\cite{Valle:1987gv,Roulet:1991sm,Guzzo:1991hi,Grossman:1995wx,Nunokawa:1996ve,Nunokawa:1996tg,Bergmann:1997mr,Krastev:1997cp}
in the early stage (before the experimental discovery of neutrino
oscillation~\cite{Fukuda:1998mi}) and afterwards, studied by a large
number of authors, for example, in
\cite{GonzalezGarcia:1998hj,Bergmann:1998rg,Bergmann:1999rz,Lipari:1999vh,Guzzo:2000kx,GonzalezGarcia:2001mp,Gago:2001xg,Berezhiani:2001rs,Huber:2001zw,Fornengo:2001pm,Huber:2002bi,Gago:2001si,Berezhiani:2001rt,Ota:2001pw,Davidson:2003ha,Friedland:2004ah,GonzalezGarcia:2004wg,Guzzo:2004ue,Friedland:2005vy,Kitazawa:2006iq,Mangano:2006ar,Blennow:2005qj,EstebanPretel:2007yu,Barranco:2007tz,Ribeiro:2007ud,Kopp:2007mi,Blennow:2008ym,EstebanPretel:2008qi,Barranco:2008,Kopp:2008ds,Kopp:2007ne,Blennow:2007pu,Ohlsson:2008gx,Gavela:2008ra,Bolanos:2008km,Antusch:2008tz,Biggio:2009nt,Escrihuela:2009up,Gago:2009ij,Biggio:2009kv,Oki:2010uc,Kopp:2010qt,Forero:2011zz,Escrihuela:2011cf,GonzalezGarcia:2011my,Coloma:2011rq,Adhikari:2012vc,Agarwalla:2012wf,Gonzalez-Garcia:2013usa,Ohlsson:2013epa,Ohlsson:2012kf,Ohlsson:2013nna,Girardi:2014kca,Agarwalla:2014bsa}.
In this short review, 
the current status of NSI is discussed, mainly 
from the phenomenological point of view.

Phenomenologically, NSI can be described with an effective four
fermion Lagrangian~\cite{Davidson:2003ha},
\begin{equation}
-{\cal L}^{eff}_{\rm NSI} =
\varepsilon_{\alpha \beta}^{fP}{2\sqrt2 G_F} (\bar{\nu}_\alpha \gamma_\rho L \nu_\beta)
( \bar {f} \gamma^\rho P f ) \,,
\label{eq:efflag}
\end{equation}
where $G_F$ is the Fermi constant, $\varepsilon_{\alpha \beta}^{fP}$
is the parameter which describes the strength of the NSI, $f$ is a
first generation SM fermion ($e, u$ or $d$), $P$ denotes the chiral
projector $\{L,R=(1\pm\gamma^5)/2\}$, and $\alpha$ and $\beta$ denote
the neutrino flavors: $e$, $\mu$ or $\tau$.
This Lagrangian describes neutral current (NC) interactions and they
will be the focus of this work.

While the Lagrangian in (\ref{eq:efflag}) provides the general
description of NSI, it is also possible to study NSI using other
approaches, defining the parameters depending on whether we are
considering the neutrino at the production point, $\epsilon^S$, during
propagation (taking into account matter effects) $\epsilon^m$, 
or at the detection point, $\epsilon^D$.  
The reader must be aware that there is no universal notation for the
NSI parameters, and some authors use similar notation for different
quantities, although the definition of the parameters in terms of
Eqs. (\ref{eq:efflag}) and~(\ref{eq:effcc}) (see below) is more
standard.

Let us consider the presence of NSI in the neutrino source with more
detail.  In general, source neutrino fluxes are produced from charged
current (CC) interactions such as pion and muon decay and, in the
presence of additional non-standard CC interactions, it would be
necessary to consider contributions that would include terms
proportional to
\begin{equation}
2{G_F}\sum_\alpha \varepsilon^{CC}_{l\alpha}
[\bar{l}(1-\gamma_5)\gamma^\rho\nu_\alpha].
\label{eq:effcc}
\end{equation}
In this case, the experimental value of the Fermi constant will be
given by~\cite{Huber:2002bi}
$G^\text{exp}=G_F 
\sqrt{ 
|1+\varepsilon^{CC}_{ee}|^2+|\varepsilon^{CC}_\mu|^2
+|\varepsilon^{CC}_\tau|^2 }
$ 
and would imply that the new interactions 
could be parameterized as~\cite{GonzalezGarcia:2001mp}
\begin{equation}
\epsilon^S_{e\alpha} = 
\frac{\varepsilon^{CC}_{e\alpha}}
{
\sqrt{|1+\varepsilon^{CC}_{ee}|^2+|\varepsilon^{CC}_\mu|^2
+|\varepsilon^{CC}_\tau|^2}  
}. 
\end{equation}
A similar expression could be obtained for the case of neutrino
detection $\epsilon^D$ if we are interested in 
CC interactions. 
It is important to notice that for NC NSI the corresponding expression
for $\epsilon^{S,D}$ is different. 
As will be discussed
below, for NSI coming from NC, the left and right couplings
$\epsilon^{L,R}$ appears naturally, while 
the $\varepsilon^{CC}_{\alpha\beta}$ for the CC case 
are considered to be left handed.

For the case of neutrino propagation, there is 
a direct relation between $\epsilon^m$ and the NSI parameters coming from 
the Lagrangian in Eq.~(\ref{eq:efflag}). It will be seen in  
section~\ref{sec:propagation} that, during propagation in matter, the 
neutrino potential will be sensitive only to vector currents 
($\epsilon^V=\epsilon^L+\epsilon^R$) and, therefore, 
we will end up with the relation,
\begin{equation}
\epsilon^m_{\alpha\beta} = \sum_f{\frac{V_f}{V_e} 
  (\varepsilon^{fL}_{\alpha\beta} + \varepsilon^{fR}_{\alpha\beta}) },
\end{equation}
where $V_f = V_f(r) \equiv \sqrt 2 G_F N_f(r)$ with $f=e, u$ or $d$
(see section 3).

Solar, atmospheric and long baseline neutrino oscillation experiments
are expected to give better constraints on propagation NSI parameters
coming from matter effects, while non oscillation experiments are more
sensitive to NSI in production and/or detection.  Both types of 
experiments provide valuable complementary information on NSI.
One of the main disadvantages of non-oscillation experiments is that the flavor
changing NSI will be present in the interaction only at the second
order level ($\varepsilon^2$) while propagation effects appear at
first order ($\varepsilon$). 
However, non-oscillation experiments also
have some advantage, for instance, they may also be sensitive to axial
currents, while oscillation experiments are not.

A model independent analysis that considers all the contributions
coming from Eq.~(\ref{eq:efflag}) will imply a large number of free
parameters, $\varepsilon^{fL}_{\alpha\beta}$.  To our knowledge, an
analysis considering all the NSI contributions at the same time has
never been done. In practice, one must constrain the analysis to a
number of parameters that could be handled by current computational
methods and give useful information about the freedom for new physics
in the neutrino sector.

In this work we will discuss the current status of NSI studies.
We will start
by giving some examples, in section~\ref{sec:models}, of the kind of
new physics that can be tested by using this formalism.  Although we
will stress the case of NC interaction, the result could be converted
into constraints for new physics in the CC sector.  We will show the
current constraints coming from both oscillation and non-oscillation
experiments in sections~\ref{sec:propagation} and~~\ref{sec:detection}
along with a brief explanation of the phenomenological procedure to
obtain such bounds.  Future perspectives to improve the current
constraints will be discussed in section~\ref{sec:future}. Finally,
conclusions will be given in section~\ref{sec:conclusions}.

\section{NSI and models for new physics}
\label{sec:models}

Although the NSI formalism appears as a correction to the vector and
axial couplings, it can account for different types
of new physics. In this section, three different classes of Standard
Model extensions are described, in term of the NSI parameters, and a
particular example is shown in every case as an illustration of the
formalism.

\subsection{Extended gauge symmetries} 

Any extension of the SM local gauge symmetry $SU(2)_L\otimes U(1)_Y$,
in general, introduces new gauge bosons that modify the vector and
axial coupling.  Typical examples are $E_6$ string inspired models,
that at low energies introduce the extra groups $U(1)_\chi\otimes
U(1)_\psi$, leading to an additional $Z'$ neutral gauge bosons. A
similar situation happens with the left-right symmetric model
$SU(2)_L\otimes U(1)_Y\otimes SU(2)_R\otimes U(1)_Y'$ where one extra
neutral and one extra charged gauge bosons, $Z'$ and $W'$, appear.

Within the NSI formalism, there is an easy direct relation between the
phenomenological parameters and the parameters coming from these
models. We can consider, for instance, the neutrino dispersion off
nuclei, described in  the SM Lagrangian
\begin{eqnarray}
\label{lagrangian}
\hspace{-1.8cm}
{\cal L}_{\nu N}^{NC}=-\frac{ G_F}{\sqrt{2}}
\sum_{{q=u,d}}
\left[\bar \nu_e \gamma^\mu (1-\gamma^{5} ) \nu_e \right] 
\left\{ f^{qL}\left[\bar q\gamma_\mu (1-\gamma^{5}) q\right] +
f^{qR}\left[\bar q\gamma_\mu (1+\gamma^{5}) q\right] \right\},
\end{eqnarray}
where $f^{qL,R}$ are the SM coupling constants defined
elsewhere~\cite{Agashe:2014kda}.  For the case of $E_6$ models, where
two additional neutral vector bosons arise, there will be an
additional contribution to these couplings constants given
by~\cite{Barranco:2007tz}
\begin{eqnarray}
\hspace{-1.5cm} \varepsilon^{uL} & = & 
- 4 \frac{M_Z^2}{M_{Z'}^2} \sin^{2}\theta_{W} \rho_{\nu N}^{NC}
\left({\cos\beta \over \sqrt{24}}-{\sin\beta \over 3}\sqrt{5 \over 8} \right)
\left({3 \cos\beta \over 2 \sqrt{24}}+{\sin\beta \over 6}\sqrt{5 \over 8} \right)
\nonumber \\
\hspace{-1.5cm} \varepsilon^{dR} &=& -8 \frac{M_Z^2}{M_{Z'}^2}
\sin^2\theta_W \rho_{\nu N}^{NC}
\left({3 \cos\beta \over 2 \sqrt{24}}+
{\sin\beta \over 6}\sqrt{5 \over 8} \right)^2 ,\nonumber \\
\hspace{-1.5cm} \varepsilon^{dL} &=& \varepsilon^{uL} = -\varepsilon^{uR}, 
\end{eqnarray}
where $M_Z$ is the mass of the SM neutral gauge boson; $M_{Z'}$
accounts for the mass of an additional, heavier, new gauge boson;
$\theta_W$ is the weak mixing angle; $\rho_{\nu N}^{NC}$ is the
parameter which accounts for the radiative corrections; and the angle
$\beta$ describes the mixing between the two extra gauge bosons that
arise from the $U(1)_\chi$ and $U(1)_\psi$ symmetries. These models
have yet another extra gauge boson that is considered to be heavier
than the $Z'$ and decoupled from the above Lagrangian.

\subsection{Additional neutral leptons}

An extension of the SM fermion content can give rise to a rich
phenomenology, especially when it contains extra neutral leptons,
usually assumed to be heavy. Within this framework, it is possible to
work in the standard $SU(2) \otimes U(1)$ gauge symmetry and consider
the additional mixing of isodoublet and isosinglet neutral
leptons~\cite{schechter:1980gr}. As a result, the $V-A$ couplings will
deviate from the SM prediction.  For example, for the case of the CC,
the ordinary light isodoublet neutrinos will mix with the extra heavy
isosinglets; this mixing will be described by a matrix
\begin{equation}
    K = ( K_L, K_H ),
    \label{eq:MCC}
\end{equation}
where $K_L$ and $K_H$ describe, respectively, 
the mixing of ordinary isodoublet light neutrinos 
and extra heavy isosinglets.
Notice that the signals of new physics coming from this matrix might
appear in charged currents through the deviations from the standard
interactions (that is, in the detection) as well as in oscillation
experiments, since the mixing matrix $K_L$ is no longer
unitary. Moreover, regarding the NC interactions, they will be
described by the interaction
\begin{equation}
    \mathcal{L} = \frac{ig^\prime}{2 \sin\theta_{W}} Z_{\mu} \bar{\nu_{L}} 
      \gamma_{\mu} K^{\dagger} K \nu_L \, , 
    \label{eq:MNC}
\end{equation}
where the matrix  $K^\dagger K$ can be  considered as a natural source for 
NSI in the neutral sector~\cite{schechter:1980gr}. 

The enriched structure that arise from these models can be parameterized
in different forms that must take into account all the new mixing
angles and phases. One of the most studied schemes in this context is
the seesaw
model~\cite{Minkowski:1977sc,gell-mann:1980vs,yanagida:1979,mohapatra:1980ia,schechter:1980gr},
which gives a natural explanation for the smallness of the neutrino
mass. In these models, however, the sizeable signals at low
energies are expected to be negligible and, therefore, the effective
NSI should be negligible. There are, however, other models where the
low energy effects, though small, may be sizeable in the near future,
such as the so-called inverse
seesaw model~\cite{mohapatra:1986bd,Forero:2011pc}.

Although it is not common in the literature to consider these analyses
in term of the NSI formalism, it is possible to include them in the
formalism. For example, for the simple case of only one extra neutral
heavy lepton, the effects on a neutrino electron scattering off
electrons will be a global factor in the Lagrangian, due to the
non-unitarity of the mixing matrix. In this case the corresponding NSI
parameters for an electron neutrino experiment will be given as
\begin{eqnarray}
\varepsilon^{eL}_{ee} = - g_L \sin^2\theta_{14}, \  \ 
\varepsilon^{eR}_{ee} = - g_R \sin^2\theta_{14}, 
\end{eqnarray}
where $\theta_{14}$ is the mixing angle between the light neutrino and
the extra heavy fermion and $g_{L,R}$ are the SM coupling constants
for the neutrino electron scattering process.

\subsection{Additional scalars}

Despite the NSI formalism preserves the $V-A$ structure of the theory, 
it is also possible to consider the impact of 
scalar couplings. For example, if we consider the case of 
low energy supersymmetry with broken R-parity~\cite{Hall:1984id,Ross:1985yg,santamaria:1987uq} where one has 
trilinear $L$ violating couplings of the form 
\begin{eqnarray}
\lambda_{ijk}L_iL_jE_k^c,  \, \ \ 
 \lambda'_{ijk}L_iQ_jD_k^c
\end{eqnarray}
with $L$ and $Q$ super-fields that contain
the usual lepton and quark $SU(2)$ doublets,
$E^c$, and $D^c$ super-fields that contain the singlets, 
and $i,j,k$ the generation indices.
These couplings give rise, for example, to the following four-fermion
effective Lagrangian for neutrino interactions with $d$-quark
\begin{equation}
   \mathcal{L}_\mathrm{eff}  =  -2\sqrt2 G_{F} \sum_{\alpha,\beta}
    \varepsilon^{dR}_{\alpha\beta}  \bar{\nu}_{L\alpha} \gamma^{\mu} \nu_{L\beta} 
    \bar{d}_{R}\gamma^{\mu}{d}_{R},
\end{equation}
where we have, among others, 
flavor-conserving and changing NSI, given, respectively, by  
\begin{eqnarray}
 \varepsilon^{dR}_{\mu\mu} &= \sum_j \frac{|\lambda'_{2j1}|^2}
{4 \sqrt{2} G_F m^2_{\tilde{q}_{j L}}} \, \  \ \text{and} \ \ 
   \varepsilon^{dR}_{\mu\tau} =  \sum_j \frac{ \lambda^\prime_{3j1} \lambda^\prime_{2j1} }
    {4\sqrt{2}G_F m^2_{\tilde{q}_{jL}} } .
\end{eqnarray}
Here,  $m_{\tilde{q}_{j L}}$ denotes  the masses of the  squarks, while $j =
1,2,3$ stands for $\tilde{d}_L, \tilde{s}_L, \tilde{b}_L$, respectively.
This is just one example, but other NSI couplings can be studied in
this context~\cite{Sessolo:2009ug}; moreover, constraints on generic
scalar NSI can also be studied by using Fierz
transformations~\cite{Gaitan:2013fma}

\section{NSI phenomenology in propagation}
\label{sec:propagation}
In this section, we discuss the phenomenological impact of 
the presence of NSI in propagation. The effect of NSI 
can be present through the modification of the 
matter potential that can exist not only in the diagonal 
but also in the off-diagonal elements in the effective 
Hamiltonian. Before starting our discussion on NSI we will briefly  review 
the current standard oscillation status. 

\subsection{Standard neutrino oscillations picture }

Unless otherwise stated, the standard 
three-flavor picture of neutrinos
is assumed, $\nu_e$, $\nu_\mu$ and $\nu_\tau$ 
and corresponding anti-particles.
In vacuum, the mixing of neutrinos is supposed to be described by the
usual flavor mixing without NSI,
\begin{eqnarray} 
|\nu_\alpha \rangle  
= \sum_{i=1}^3 U^\ast_{\alpha i}\ |\nu_i  \rangle\ 
\ (\alpha = e, \mu, \tau), 
\end{eqnarray} 
where $U$ is the 3 $\times$ 3 matrix which describes 
the flavor mixing ~\cite{Maki:1962mu} of neutrinos.
In this review, we use the standard parameterization found, e.g, 
in \cite{Agashe:2014kda}, 
\begin{eqnarray}
\hskip -1.5cm 
U &=& 
\left[
\begin{array}{ccc}
1  &  0 & 0      \cr 
0 &  c_{23}& s_{23} \cr 
0 & -s_{23}  & c_{23}  \cr 
\end{array}
\right]
\left[
\begin{array}{ccc}
c_{13}  &  0 &     s_{13}e^{-i\delta_{CP}}  \cr 
0       &   1&  0 \cr 
-s_{13}e^{i\delta_{CP}} & 0  & c_{13}  \cr 
\end{array}
\right]
\left[
\begin{array}{ccc}
c_{12}  &  s_{12} &  0  \cr 
-s_{12} &  c_{12} &  0 \cr 
0  & 0  & 1  \cr 
\end{array}
\right], 
\nonumber \\
\hskip -1.5cm &=& 
\left[
\begin{array}{ccc}
c_{12}c_{13}  &  s_{12}c_{13} & s_{13} e^{-i\delta_{CP}}      \cr 
-s_{12}c_{23} -c_{12}s_{23}s_{13} e^{i\delta_{CP}}  
&  c_{12}c_{23} -s_{12}s_{23}s_{13} e^{i\delta_{CP}}  
& s_{23} c_{13} \cr 
s_{12} s_{23} -c_{12}c_{23}s_{13} e^{i\delta_{CP}} &
-c_{12} s_{23} -s_{12}c_{23}s_{13} e^{i\delta_{CP}} &
c_{23}c_{13}  \cr 
\end{array}
\right],
\label{MNS-matrix}
\end{eqnarray}
where $s_{ij} \equiv \sin \theta_{ij}$, 
$c_{ij} \equiv \cos \theta_{ij}$, and 
$\delta_{CP}$ is the Kobayashi-Maskawa~\cite{Kobayashi:1973fv} 
type CP phase for neutrinos.

In addition to the mixing angles and CP phase, 
the mass squared differences of neutrinos, $\Delta m^2_{ij} \equiv
m_i^2-m_j^2$ with $m_{i}$ ($i$=1-3) being the neutrino masses, 
are the relevant parameters to describe neutrino oscillation.
So far, all of these parameters have been measured with reasonably 
good accuracies except for the value of the CP phase 
and sign of $\Delta m^2_{31}$ ($\Delta m^2_{32}$). 
The positive (negative) sign of $\Delta m^2_{31}$ 
corresponds to the normal  (inverted) mass ordering,  
often referred to as normal (inverted) mass hierarchy.

Different groups have carefully studied the neutrino data and obtained
accurate values for most of the three-flavor neutrino oscillation
parameters~\cite{Capozzi:2013csa,Forero:2014bxa,Gonzalez-Garcia:2014bfa}.
Their most important results are summarized in
Table~\ref{tab:Table-standard}, where a reasonable agreement can be
seen.

\begin{table}[!h]
\centering
\caption{Summary of the standard three-flavor picture parameters, as
  reported from three different groups, denoted as C (Capozzi el
  al~\cite{Capozzi:2013csa}, second and third column, F (Forero et
  al~\cite{Forero:2014bxa}, fourth and fifth column), and 
G  (Gonzalez-Garcia et al~\cite{Gonzalez-Garcia:2014bfa}, the last
 two columns).  
The  parameter $\Delta m^2_{3l}$ 
has a slightly different definition in
each case, being $\Delta m^2_{3l} \equiv m^2_3 - (m^2_1 + m^2_2)/2$ for
  Ref.~\cite{Capozzi:2013csa}, $\Delta m^2_{3l} \equiv m^2_3-m^2_1$ for
  Ref.~\cite{Forero:2014bxa} and $\Delta m^2_{3l} \equiv m^2_3-m^2_1$ for
  normal hierarchy (NH) and $\Delta m^2_{3l} \equiv m^2_3-m^2_2$ for
  inverted hierarchy (IH) for Ref.~\cite{Gonzalez-Garcia:2014bfa} }
\vskip .2cm
\begin{tabular}{ccccccc}
\hline\hline
 & \multicolumn{2}{c}{C~\cite{Capozzi:2013csa}} &
       \multicolumn{2}{c}{F~\cite{Forero:2014bxa}} & 
       \multicolumn{2}{c}{G~\cite{Gonzalez-Garcia:2014bfa}} \\ 
Parameter & Best fit & $3\sigma$ range & Best fit & $3\sigma$ range & 
                                     Best fit & $3\sigma$ range\\ 
\hline
$\Delta m^2_{21}/10^{-5}$~eV$^2$ & 7.54 & 6.99-8.18 & 7.60 & 7.11-8.18
                                          & 7.50 & 7.02-8.09 \\ 
$\Delta m^2_{3l}/10^{-3}$~eV$^2$~(NH) & 2.43 & 2.23-2.61 & 2.48 & 2.30-2.65 & 
                                           2.457& 2.317-2.607 \\ 
$-\Delta m^2_{3l}/10^{-3}$~eV$^2$~(IH) & 2.38 & 2.19-2.56 & 2.38 & 2.20-2.54 &
                                           2.449 & 2.307-2.590 \\ 
$\sin^2\theta_{12}/10^{-1}$ & 3.08 & 2.59-3.59 & 3.23 & 2.78-3.75 & 
                                           3.04 & 2.70-3.44 \\ 
$\sin^2\theta_{23}/10^{-1}$~(NH) & 4.37 & 3.74-6.26 & 5.67 & 3.93-6.43 & 
                                           4.52 & 3.82-6.43 \\ 
$\sin^2\theta_{23}/10^{-1}$~(IH) &  4.55 &
	 3.80-6.41 
& 5.73 & 4.03-6.40 & 
                                           5.79 & 3.89-6.44  \\ 
$\sin^2\theta_{13}/10^{-2}$~(NH) & 2.34 & 1.76-2.95 & 2.26
& 1.90-2.62 &
                                             2.18 & 1.86-2.50 \\ 
$\sin^2\theta_{13}/10^{-2}$~(IH) & 2.40 & 1.78-2.98
& 2.29 & 1.93-2.65 & 
                                             2.19 & 1.88-2.51  \\ 
$\delta/^o$~(NH) & 250 & 0-360 & 254 & 0-360 & 306 & 0-360  \\ 
$\delta/^o$~(IH) & 236 & 0-360 & 266 & 0-360 & 254 & 0-360 \\
\hline\hline
\end{tabular}
\label{tab:Table-standard}
\end{table}

\subsection{Neutrino evolution with NSI}

Phenomenologically, the evolution equation of neutrinos 
in the flavor basis in the presence of propagation 
NSI in unpolarized matter 
can be generically written as,  
\begin{eqnarray} 
\hspace{-2.7cm}
i {d\over dr} \left[ \begin{array}{c} 
                   \nu_e \\ \nu_\mu \\ \nu_\tau 
                   \end{array}  \right]
=  
\left\{
U \left[ \begin{array}{ccc}
       0   & 0          & 0   \\
       0   &\Delta_{21} & 0  \\
       0   & 0          &\Delta_{31}  
\end{array} \right]U^{\dagger} 
 + \sum_f V_f
\left[ \begin{array}{ccc}
\delta_{ef}+ \varepsilon_{ee}^f 
& \varepsilon_{e\mu}^f & \varepsilon_{e\tau}^f \\
{\varepsilon^f}_{e \mu }^*  & \varepsilon_{\mu\mu}^f  
& \varepsilon_{\mu\tau}^f \\
{\varepsilon^f}_{e \tau}^* & {\varepsilon^f}_{\mu \tau }^* & \varepsilon_{\tau\tau}^f 
\end{array} 
\right] 
\right\}
\left[\begin{array}{c} 
                   \nu_e \\ \nu_\mu \\ \nu_\tau 
                   \end{array}\right], 
\label{eq:general-evolution}
\end{eqnarray}
where $\nu_\alpha \equiv \langle \nu_\alpha | \nu(r)\rangle \ (\alpha
= e, \mu, \tau)$ denotes the probability amplitude to find neutrino as
$\nu_\alpha$ at the position $r$, $\Delta_{ij} \equiv \Delta
m^2_{ij}/(2E)$, $E$ being the neutrino energy.
$V_f = V_f(r) \equiv \sqrt 2 G_F N_f(r) $ where $N_f$ ($f=e,
u$ or $d$) denotes the fermion number density along the neutrino
trajectory in matter.  Note that $V_e(r)$ is the standard matter
potential \cite{Wolfenstein:1977ue} which induces the usual MSW
effect~\cite{Mikheev:1986gs,Wolfenstein:1977ue}.

Since NSI effects in propagation enter only through the vector
couplings, $\varepsilon_{\alpha\beta}^f$ must be interpreted as
$\varepsilon_{\alpha\beta}^f = \varepsilon_{\alpha\beta}^{fL} +
\varepsilon_{\alpha\beta}^{fR}$.  For simplicity, throughout this
review, we consider the case where only $d$-quark has the propagation
NSI with neutrinos, and write NSI parameters simply as
$\varepsilon_{\alpha\beta}$ by omitting the fermion superscript.  Note
that the case of $u$-quark NSI is very similar in most cases to be
discussed in this section because in the usual matter, $N_u \sim N_d
\sim 3 N_e$.

In this review, unless otherwise stated, 
for definiteness, we use the following values of 
the mixing parameters as our reference values; 
$\sin^2 \theta_{12} = 0.31$, $\sin^2 \theta_{13} = 0.023$, 
$\sin^2 \theta_{23} = 0.5$, 
$\Delta m^2_{21}  = 7.5 \times 10^{-5}$ eV$^2$ and
$|\Delta m^2_{31}| = 2.4 \times 10^{-3}$ eV$^2$,
and $\delta_{CP} = 0$, 
which are consistent at 2$\sigma$ with 
the results obtained by the recent global 
analysis~\cite{Capozzi:2013csa,Forero:2014bxa,Gonzalez-Garcia:2014bfa}. 

Eq.~(\ref{eq:general-evolution}) defines the framework for neutrino
propagation in matter with NSI.
The parameters $\varepsilon_{\alpha\beta}$ ($\alpha, \beta = e, \mu, \tau$)
describe the magnitude of NSI. 
The diagonal NSI parameters, $\varepsilon_{\alpha\alpha}
(\alpha=e,\mu,\tau)$, could play a role similar to the terms of the
standard MSW matter potential, or could be interpreted as the NSI
induced mass squared difference, mimicking the ones that contain
$\Delta m^2$, which could induce new resonance even if neutrinos were
massless \cite{Valle:1987gv,Guzzo:1991hi}.
On the other hand, off-diagonal NSI parameters,  
$\varepsilon_{\alpha\beta} (\alpha \ne \beta)$ could 
play a role similar to the mixing angle. 
Even if there is no mixing in vacuum, 
the flavor transitions $\nu_\alpha \to \nu_\beta$ can 
occur in matter due to the presence of 
the off-diagonal NSI~\cite{Valle:1987gv,Roulet:1991sm,Guzzo:1991hi}.
The complex phases of the off-diagonal elements
$\varepsilon_{\alpha\beta}$ could be a new source of 
CP violation, see e.g., \cite{GonzalezGarcia:2001mp,Gago:2009ij}. 

Currently, almost all the neutrino data are consistent with the
standard three flavor scheme of massive and mixed neutrinos. 
Therefore, NSI, if they exist, is expected to manifest only as a
subdominant effect.
NSI in propagation has been constrained mainly by the oscillation data
of solar and atmospheric neutrinos as well as neutrinos produced by
accelerators.  Reactor neutrinos do not constrain the propagation NSI
because the matter effect is expected to be very small though they can
constrain the detection NSI.

Roughly speaking, for a given neutrino energy, and the matter density,
$\rho$, the impact of propagation NSI in neutrino oscillation
(modification of the standard oscillation due to NSI) is essentially
determined by the magnitude of the following dimensionless quantity,
\begin{eqnarray}
\eta_{\alpha\beta} \equiv \frac{\varepsilon_{\alpha\beta} V_f}{\Delta_{ij}}
\approx 0.1 \times \varepsilon_{\alpha\beta} 
\left[ \frac{E}{\text{GeV}}\right]\ 
\left[ \frac{2.4\times 10^{-3}\ \text{eV}^2 }{ \Delta m^2_{ij}}\right]\ 
\left[ \frac{\rho}{\text{g/cm}^3}\right],
\end{eqnarray}
where the NSI with $d$ or $u$ quarks are assumed; $\Delta m^2_{ij}$ is
the relevant mass squared difference in the corresponding oscillation
channel,  and the baseline $L$ is assumed to be large
enough, or $\Delta_{ij} L \gsim O(1)$.
Larger the value of $\eta_{\alpha\beta}$, 
larger the impact of NSI in propagation. 

\subsection{NSI for atmospheric neutrinos}

In this subsection we discuss the NSI effect for atmospheric neutrinos. 
The impact of NSI on atmospheric neutrinos have been
considered by many authors, see e.g., 
~\cite{GonzalezGarcia:1998hj,Lipari:1999vh,Fornengo:2001pm,
GonzalezGarcia:2004wg, 
Friedland:2004ah,Friedland:2005vy,GonzalezGarcia:2011my}.

Let us first consider the impact of NSI on 
the $\nu_\mu-\nu_\tau$ sector and assume that
all the NSI parameters coupling to electron flavor neutrino, namely, 
$\varepsilon_{e \beta}$ are zero. 
In this scenario, 
$\varepsilon_{\mu \tau}$ and 
$\varepsilon_{\tau\tau}-\varepsilon_{\mu\mu}$ can be constrained
mainly by the higher energy samples of the 
atmospheric neutrino data as will be seen below. 

In the limit of $\Delta m^2_{21}L/E \to 0$, 
with the constant matter density approximation, 
and ignoring $\theta_{13}$, the $\nu_\mu \to \nu_\mu$ 
survival probability is expressed 
as~\cite{Gago:2001xg,GonzalezGarcia:2004wg}
\begin{eqnarray} 
P(\nu_\mu \to \nu_\mu) 
= 
1-P(\nu_\mu \to \nu_\tau) 
\simeq  1- \sin^2 2\theta_{\text{eff}} 
\sin^2 \left[ \xi \ \frac{\Delta_{31} L}{2} \right], 
\end{eqnarray}
where 
\begin{eqnarray} 
\sin^2 2\theta_{\text{eff}} \equiv 
\frac{ | \sin 2\theta_{23} \pm  2\eta_{\mu\tau}|^2 }
{\xi^2}, 
\end{eqnarray}
\begin{eqnarray}
\xi \equiv 
\sqrt{ 
  | \sin 2\theta_{23} \pm  2\eta_{\mu\tau} |^2
+ \{\cos 2\theta_{23} + (\eta_{\mu\mu}-\eta_{\tau\tau}) \}^2
},
\end{eqnarray}
and the $+(-)$ sign in front of $\eta_{\mu\tau}$ corresponds
to the normal (inverted)  mass ordering. 
For anti-neutrino channel, the sign of $\eta_{\alpha\beta}$
must be changed. 

\begin{figure}[!h]
\vglue -1.8cm
\begin{center}
\includegraphics[width= 0.92\textwidth]{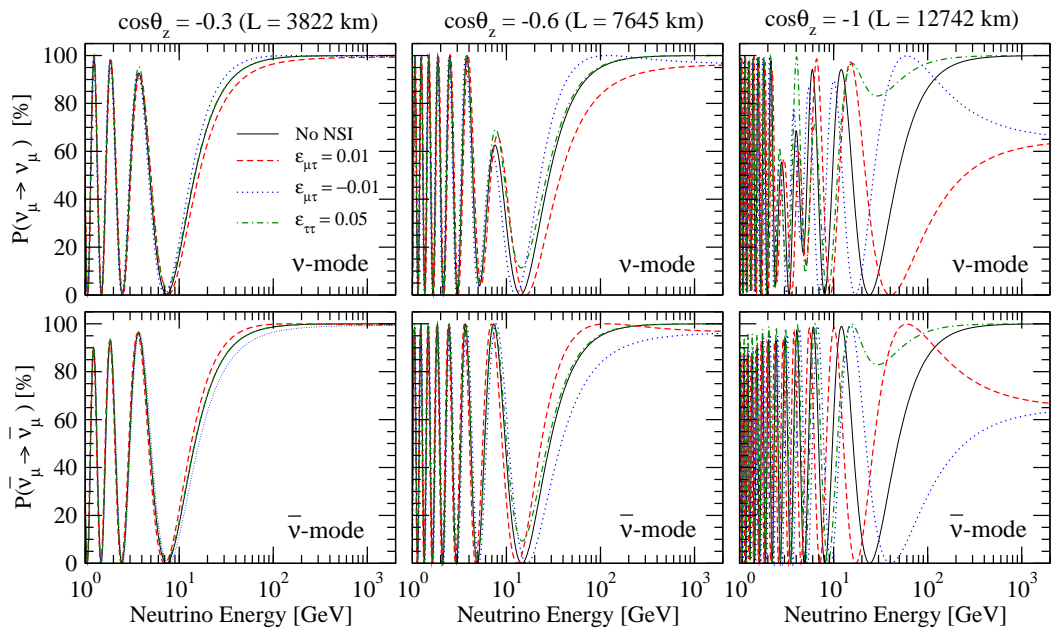}
\end{center}
\vglue -1.7cm
\caption{
Survival probabilities of $\nu_\mu\to\nu_\mu$ (upper panels)
and $\bar{\nu}_\mu\to\bar{\nu}_\mu$ (lower panels) as 
a function of the neutrino energy for 
different zenith angle of incoming neutrinos, 
$\cos \theta_z = -0.3$ (left panels), $-0.6$ (middle panels) 
and $-1$ (right panels). The corresponding distances traveled by 
neutrinos are indicated in the plots. The normal mass 
ordering was assumed. 
}
\label{fig:Pmm_atm_ene_NH}
\vglue -0.2cm
\end{figure}

For the atmospheric neutrinos, the sensitivity to the NSI parameters
can be estimated by studying the muon neutrino and anti-neutrino
survival probabilities for different zenith angles, $\cos \theta_z$.
Fig.~\ref{fig:Pmm_atm_ene_NH} shows the muon neutrino survival
probabilities (for the normal mass ordering) for $\cos \theta_z = -0.3$
(left panels), $-0.6$ (middle panels) and $-1$ (right panels) for the
cases without NSI (by solid lines) and with NSI (by non-solid lines).
For this calculation, the neutrino evolution
equation~(\ref{eq:general-evolution}) was solved numerically 
(without ignoring neither $\Delta m^2_{21}$ nor $\theta_{13}$) 
using the Earth matter density profile predicted in 
the Preliminary Reference Earth Model (PREM) model~\cite{Dziewonski:1981xy}.

Note that, by comparing the upper and lower panels, 
the dependence on the sign of NSI parameters for 
neutrinos and anti-neutrinos are opposite. 
Note also that, with a good approximation,
the survival probabilities are invariant under
the simultaneous 
transformation $\Delta_{31} \to -\Delta_{31}$
and $\varepsilon_{\mu\tau} \to - \varepsilon_{\mu\tau}$. 
In the limit of the 2 flavor approximation, what is relevant
is only the relative sign of these quantities. 
As can be seen from Fig. ~\ref{fig:Pmm_atm_ene_NH}, the impact of NSI
is small for lower energies, for $E \lsim 5$ GeV, even for the case
where neutrino pass through the center of the Earth.
On the other hand, the impact of NSI for energy 
$\gsim 10$ GeV, could be quite large
for the NSI parameters considered in 
Fig. ~\ref{fig:Pmm_atm_ene_NH} and it is expected that
these values could be disfavored or excluded. 

Here we quote the bounds on these NSI parameters obtained 
by the Super-Kamiokande collaboration~\cite{Mitsuka:2011ty},  
$|\epsilon_{\mu \tau}| < 1.1 \times 10^{-2}$ and 
$-4.9 \times 10^{-2} < 
\epsilon_{\tau \tau} -\epsilon_{\mu \mu} < 4.9 \times 10^{-2}$ 
at  90\% CL. 
More recently, by using the IceCube-79 and DeepCore 
data, authors of \cite{Esmaili:2013fva} obtained somewhat 
better bounds,
$|\epsilon_{\mu \tau}| \lsim 6 \times 10^{-3}$ 
and $ |\epsilon_{\tau \tau} -\epsilon_{\mu \mu}| \lsim 3 \times 10^{-2}$
at  90\% CL.  

\begin{figure}[!h]
\vglue -1.8cm
\begin{center}
\includegraphics[width= 0.92\textwidth]{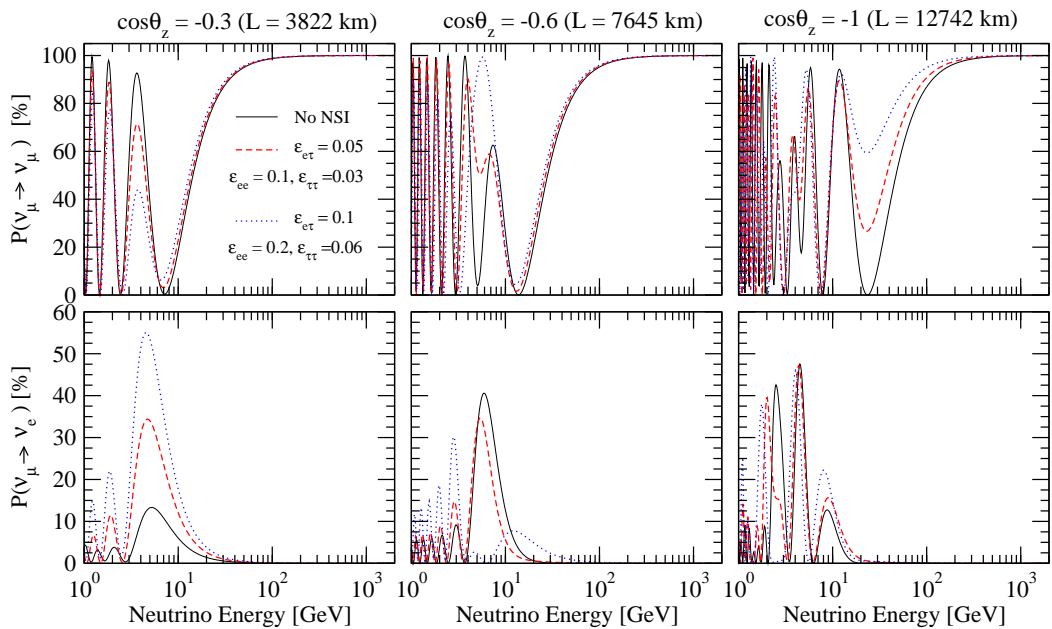}
\end{center}
\vglue -1.7cm
\caption{
Survival probabilities of $\nu_\mu\to\nu_\mu$ (upper panels)
and $\nu_\mu\to\nu_e$ (lower panels) as 
a function of the neutrino energy for 
different zenith angle of incoming neutrinos, 
$\cos \theta_z = -0.3$ (left panels), $-0.6$ (middle panels) 
and $-1$ (right panels). 
The normal mass ordering was assumed. 
}
\label{fig:Pmm_Pme_atm_ene_NH}
\vglue -0.2cm
\end{figure}

For the $\nu_e-\nu_\tau$ sector, besides the probability for $\nu_\mu
\to \nu_\mu$, it is also useful to consider the $\nu_\mu \to \nu_e$
case. The computation of these probabilities, shown in
Fig.~\ref{fig:Pmm_Pme_atm_ene_NH}, was done in an analogous way to 
the case shown in Fig.~\ref{fig:Pmm_atm_ene_NH}.  For this computation
different NSI parameters have been considered: $\varepsilon_{ee}$,
$\varepsilon_{\tau\tau}$ and $\varepsilon_{e\tau}$.  It is important
to notice that, in this case, the $\varepsilon_{e\tau}$ parameters,
could play a role similar to $\theta_{13}$. Therefore, there is some
impact on the $\nu_\mu \to \nu_e$
channel as it is possible to see 
in the lower panels of Fig.~\ref{fig:Pmm_Pme_atm_ene_NH}. 

%
%
When $\varepsilon_{ee}$, $\varepsilon_{\tau\tau}$ 
and  $\varepsilon_{e\tau}$ are assumed to be simultaneously 
nonzero, 
it is known~\cite{Friedland:2005vy} 
that the allowed combinations of the NSI parameters are 
approximately given by the parabolic relation, 
\begin{eqnarray} 
\varepsilon_{\tau\tau} \sim 
\frac{3 |\varepsilon_{e\tau}|^2} {1+3\varepsilon_{ee}}. 
\end{eqnarray} 
This feature can be also confirmed in 
Figs. 8 and 9 in ~\cite{Mitsuka:2011ty} which show
the constraints on these NSI parameters obtained by 
the Super-Kamiokande atmospheric neutrino data. 
It is also pointed out in ~\cite{Friedland:2005vy} 
that atmospheric neutrino data alone can not
essentially constrain $\varepsilon_{ee}$ parameter.
Therefore, in general one can obtain 
the allowed regions of 
$\varepsilon_{e\tau}$ and $\varepsilon_{\tau\tau}$
for given values of $\epsilon_{ee}$
as done in ~\cite{Friedland:2005vy,Mitsuka:2011ty}. 
For example, from Fig. 9 of \cite{Mitsuka:2011ty}, 
we see that for $\sin^2 \theta_{23} = 0.5$, 
for $\varepsilon_{ee} = -0.5$, 
$0$, and $0.5$, 
$\varepsilon_{e\tau} \lsim $, 
$0.08$, $0.11$ and $0.18$, respectively, 
at 90\% CL. 

\subsection{NSI for accelerator neutrinos}

\begin{figure}[!h]
\vglue -3.0cm
\begin{center}
\hglue 0.7cm
\includegraphics[width= 0.96\textwidth]{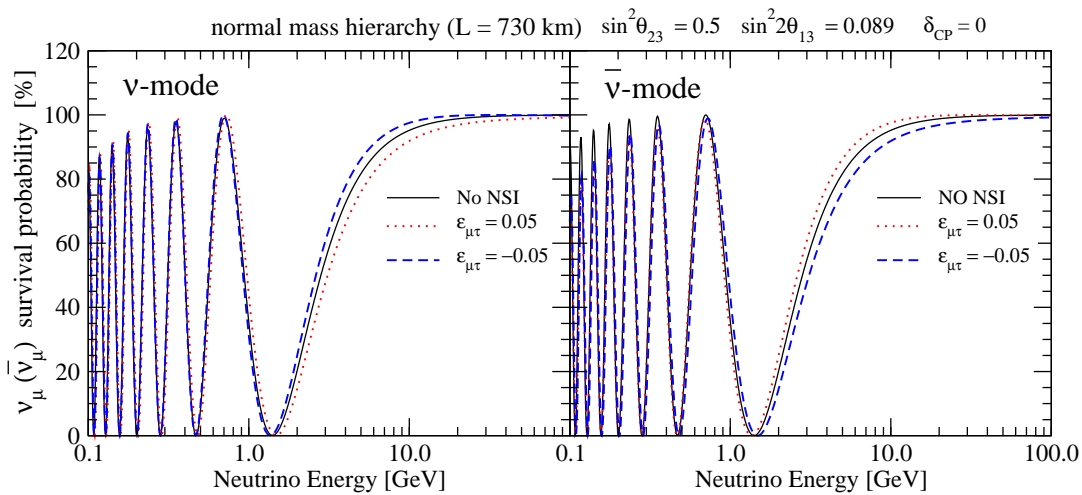}
\end{center}
\vglue -2.8cm
\caption{
$\nu_\mu \to \nu_\mu$ and 
$\bar{\nu}_\mu \to \bar{\nu}_\mu$ survival 
probabilities as a function of neutrino energy for 
the MINOS baseline, $L=730$ km without the presence of NSI 
with and with NSI, $\varepsilon_{\mu\tau} = \pm 0.1$.
The matter density was assumed to be constant, $\rho=3.2$
 g/cm$^3$. 
}
\label{fig:Pmm_energy_MINOS}
\end{figure}

So far the bounds on propagation NSI from accelerator neutrinos 
mainly come from the $\nu_\mu \to \nu_\mu$ and 
$\bar{\nu}_\mu \to \bar{\nu}_\mu$ channels. 
For these channels, at first approximation, 
the relevant NSI parameters are $\varepsilon_{\mu\tau}$, 
$\varepsilon_{\mu\mu}$ and $\varepsilon_{\tau\tau}$.
The $\nu_\mu \to \nu_\mu$ and $\bar{\nu}_\mu \to \bar{\nu}_\mu$
survival probabilities as a function of neutrino energy for the MINOS
baseline, $L=730$ km, are shown in
Fig. ~\ref{fig:Pmm_energy_MINOS}. 
The computations were done without the presence of NSI (solid lines)
and with $\varepsilon_{\mu\tau} = \pm 0.1$ 
(dotted and dashed lines).
%
As it is possible to see from Fig. ~\ref{fig:Pmm_energy_MINOS} 
the impact of NSI for $\nu$ and $\bar{\nu}$ 
channels are opposite. 

The bounds obtained by the MINOS collaboration~\cite{Adamson:2013ovz},
translated to the notation used in this review, can be stated as
$-0.067 < \varepsilon_{\mu\tau} < 0.023$ at 90\% CL.

\subsection{NSI for solar neutrinos}

The most updated analysis of solar neutrinos in the 
context of the propagation NSI comes from ~\cite{Gonzalez-Garcia:2013usa}.
For solar neutrinos, 
under the so called 
one mass scale dominance approximation, 
$|\Delta m^2_{31}| \to \infty$, 
the neutrino evolution  
can be effectively reduced to that of the
2 flavor system~\cite{Kuo:1987zx}, as follows, 
\begin{eqnarray} 
\hspace{-1.7cm}
i {d\over dr} 
\left[ \begin{array}{c} 
          \nu_e' \\ \nu_\mu' 
       \end{array}  \right]
=  
\left\{
U_{\theta_{12}}
\left[ 
\begin{array}{cc}
       0   & 0            \\
       0   &\Delta_{21}  
\end{array} 
\right]U_{\theta_{12}}^{\dagger} 
 + \sum_f V_f
\left[ \begin{array}{cc}
c^2_{13} \delta_{ef} - \varepsilon_{D}^f & \varepsilon_{N}^f \\
{\varepsilon^f}_{N}^*  & \varepsilon_{D}^f  
\end{array} 
\right] 
\right\}
 \left[\begin{array}{c} 
                   \nu_e' \\ \nu_\mu'
   \end{array}  \right],
\label{eq:solar-evolution}
\end{eqnarray}
where 
\begin{eqnarray}
\hskip -1.5cm 
\left[ 
\begin{array}{c} 
\nu_e' \\ \nu_\mu' \\ \nu_\tau' 
\end{array}  
\right]
&\equiv& 
\left[
\begin{array}{ccc}
c_{12}  &  s_{12} &  0  \cr 
-s_{12} &  c_{12} &  0 \cr 
0  & 0  & 1  \cr 
\end{array}
\right]
\left[ 
\begin{array}{c} 
\nu_1 \\ \nu_2 \\ \nu_3 
\end{array}  
\right]
= 
\left[
\begin{array}{ccc}
c_{12}  &  s_{12} &  0  \cr 
-s_{12} &  c_{12} &  0 \cr 
0  & 0  & 1  \cr 
\end{array}
\right]
U^{\dagger}
\left[ 
\begin{array}{c} 
\nu_e \\ \nu_\mu \\ \nu_\tau 
\end{array}  
\right],
\end{eqnarray}
with $\nu_\tau'$ being decoupled from the system~\cite{Kuo:1987zx}, 
and 
diagonal $\varepsilon_D$ and off-diagonal $\varepsilon_N$
NSI parameters are related to $\varepsilon_{\alpha\beta}$ 
as~\cite{Gonzalez-Garcia:2013usa},
\begin{eqnarray}
\label{eq:eps_D}
\hspace{-1.4cm}\varepsilon_D & = &
 c_{13} s_{13} \text{Re} \left[ \text{e}^{i\delta_{CP}} 
 ( s_{23} \, \varepsilon_{e\mu}
       + c_{23} \, \varepsilon_{e\tau} ) \right]
     - ( 1 + s_{13}^2 ) c_{23} s_{23} 
 \text{Re} ( \varepsilon_{\mu\tau})
\nonumber    \\
\hspace{-1.4cm}    & \hphantom{={}} &
    -\frac{c_{13}^2}{2} \big( \varepsilon_{ee} - \varepsilon_{\mu\mu} \big)
    + \frac{s_{23}^2 - s_{13}^2 c_{23}^2}{2}
    \big( \varepsilon_{\tau\tau} - \varepsilon_{\mu\mu} \big) \,, \\
\hspace{-1.4cm} \varepsilon_N 
&=&
 c_{13} \big( c_{23} \, \varepsilon_{e\mu} - s_{23} \, 
\varepsilon_{e\tau} \big)
 + s_{13} e^{-i\delta_{CP}} 
\left[   s_{23}^2 \, \varepsilon_{\mu\tau} - 
c_{23}^2 \, \varepsilon_{\mu\tau}^{\ast}
   + c_{23} s_{23} \big( \varepsilon_{\tau\tau} - \varepsilon_{\mu\mu} \big)
  \right]. \hskip 1.0cm 
\end{eqnarray}
In this approximation, the survival probability is given as
\begin{eqnarray}
P(\nu_e\to\nu_e) = s_{13}^4 + c_{13}^4 
P_{2\nu}(\nu_e'\to\nu_e'),
\end{eqnarray}
where 
$P_{2\nu}(\nu_e'\to\nu_e')$ is 
calculated for the effective 2 flavor system described 
by (\ref{eq:solar-evolution}). 
According to \cite{Gonzalez-Garcia:2013usa}, 
the bounds on these parameters are 
$-0.25 < \varepsilon_D < -0.02 $ and 
$-0.14 < \varepsilon_N < 0.12 $ at 90\% CL
assuming NSI with $d$-quark. 

In Table \ref{tab:Table-NSI-propagation}
we show the summary of the bounds on the propagation NSI. 

\begin{table}[h!]
\centering
\caption{Constraints on the matter (propagation) 
NSI parameters at 90\% C L. 
for the interaction of neutrinos with $d$ type quark.
The other NSI parameters are set to zero.}
\vskip .2cm
\begin{tabular}{ccc}
\hline\hline
NSI parameters & Bounds & Reference \\
\hline 
$  \varepsilon^{m}_{ee} -\varepsilon^{m}_{\mu\mu} $  & 
(0.02, 0.51) & \cite{Gonzalez-Garcia:2013usa} \\
$  \varepsilon^{m}_{\tau\tau} -\varepsilon^{m}_{\mu\mu} $  & 
($-0.01, 0.03$) & \cite{Gonzalez-Garcia:2013usa} \\
$  \varepsilon^{m}_{\tau\tau} -\varepsilon^{m}_{\mu\mu} $  & 
($-0.049, 0.049$)  & \cite{Mitsuka:2011ty} \\
$  \varepsilon^{m}_{\tau\tau} -\varepsilon^{m}_{\mu\mu} $  & 
($-0.036, 0.031$) & \cite{Esmaili:2013fva} \\
\hline
$  \varepsilon^{m}_{e\mu}$  & 
($-0.09, 0.04$)  & \cite{Gonzalez-Garcia:2013usa} \\
$  \varepsilon^{m}_{\mu\tau}$  & 
($-0.01$, 0.01)  & \cite{Gonzalez-Garcia:2013usa} \\
$  \varepsilon^{m}_{\mu\tau}$  & 
($-0.011, 0.011$)  & \cite{Mitsuka:2011ty} \\
$  \varepsilon^{m}_{\mu\tau}$  & 
($-6.1\times 10^{-3}$, $5.6\times 10^{-3}$) & \cite{Esmaili:2013fva} \\
$  \varepsilon^{m}_{e\tau}$  & 
($-0.13, 0.14$) & \cite{Gonzalez-Garcia:2013usa} \\
$  \varepsilon^{m}_{e\tau}$ (for $\varepsilon^{m}_{ee}=-0.50$)  & 
($-0.05, 0.05$)  & \cite{Mitsuka:2011ty} \\
$\varepsilon^{m}_{e\tau}$ (for $\varepsilon^{m}_{ee}$=0.50) & 
($-0.19, 0.13$) & \cite{Mitsuka:2011ty} \\
\hline\hline
\end{tabular}
\label{tab:Table-NSI-propagation}
\end{table}

\section{NSI phenomenology in detection}
\label{sec:detection}
Several experiments have been devoted specifically to measure with
precision the neutrino interaction with quarks and leptons.  They are
performed at very short baselines, avoiding effects coming from the
standard oscillation.  These measurements allow us to test the
validity of the interactions described by the Standard Model,
and, therefore, they could be a basis to search
for new physics beyond SM.

For non-oscillation experiments, NSI can be constrained by comparing
the measured cross sections with that predicted by SM 
for the interaction of the neutrinos with the corresponding target.
Most of these experiments record fewer events than oscillation
experiments. On the other hand, they are independent of the mixing
parameters; therefore, the cross section measurements, in general, do
not suffer from the uncertainties of the oscillation parameters.
Moreover, non-oscillation experiments are sensitive to axial
couplings, a coupling that is absent in propagation 
NSI effects.

Here we will review different experiments that constrain NSI through
detection.  
We will start by considering the
neutrino interactions with electrons and, afterwards, we will review
its interactions with quarks.

We will illustrate the phenomenology involved in this type of
experiments by considering the specific case of the electron anti-neutrino
scattering off electrons. In this case, the differential cross
section, including the corrections coming from the Lagrangian shown in 
Eq.~(\ref{eq:efflag}), will be given by 
\begin{eqnarray}
\label{eq:xsec-anu}
\hspace{-2.0cm}
\frac{{\rm d}\sigma }{{\rm d} T_e}
= \frac{2G_{F}^2 m_{e}}{\pi} 
\displaystyle \left[ 
(g_R + \varepsilon_{e e}^{R})^2 \right.
&+&\sum_{\alpha \neq e}|\varepsilon_{\alpha e}^{R}|^2 
+ \left\{ (g_L+\varepsilon_{e e}^{L})^2
          +\sum_{\alpha \neq e}|\varepsilon_{\alpha e}^{L}|^2 \right\}
          \left( 1-{T_e \over E_{\nu}}\right)^2  \nonumber \\ 
&-&
\left.  \left\{(g_L+\varepsilon_{e e}^{L})( g_R+\varepsilon_{e e}^{R}) + \sum_{\alpha \neq e}|\varepsilon_{\alpha e}^{L}||
\varepsilon_{\alpha e}^{R}| \right\}
m_e {T_e \over E^2_{\nu}} \right]. 
\end{eqnarray}
Here, $m_e$ is the electron mass, $T_e\equiv E_e - m_e$ (with $E_e$ being
the total electron energy) stands for the electron recoil energy, 
and $E_\nu$ is the anti-neutrino energy. The Standard Model couplings, at
tree level, are defined as $g_L = 1/2 + \sin^2\theta_{\rm W}$ and $g_R
= \sin^2\theta_{\rm W}$. We can see from this expression that flavor
changing NSI parameters ($\varepsilon_{\mu e}^{L,R}$ and
$\varepsilon_{\tau e}^{L,R}$) will only give quadratic corrections
while the flavor diagonal ones could give linear corrections.

\begin{figure}[!h]
\vglue -1.00cm
\centering
\includegraphics[width= 0.85\textwidth]{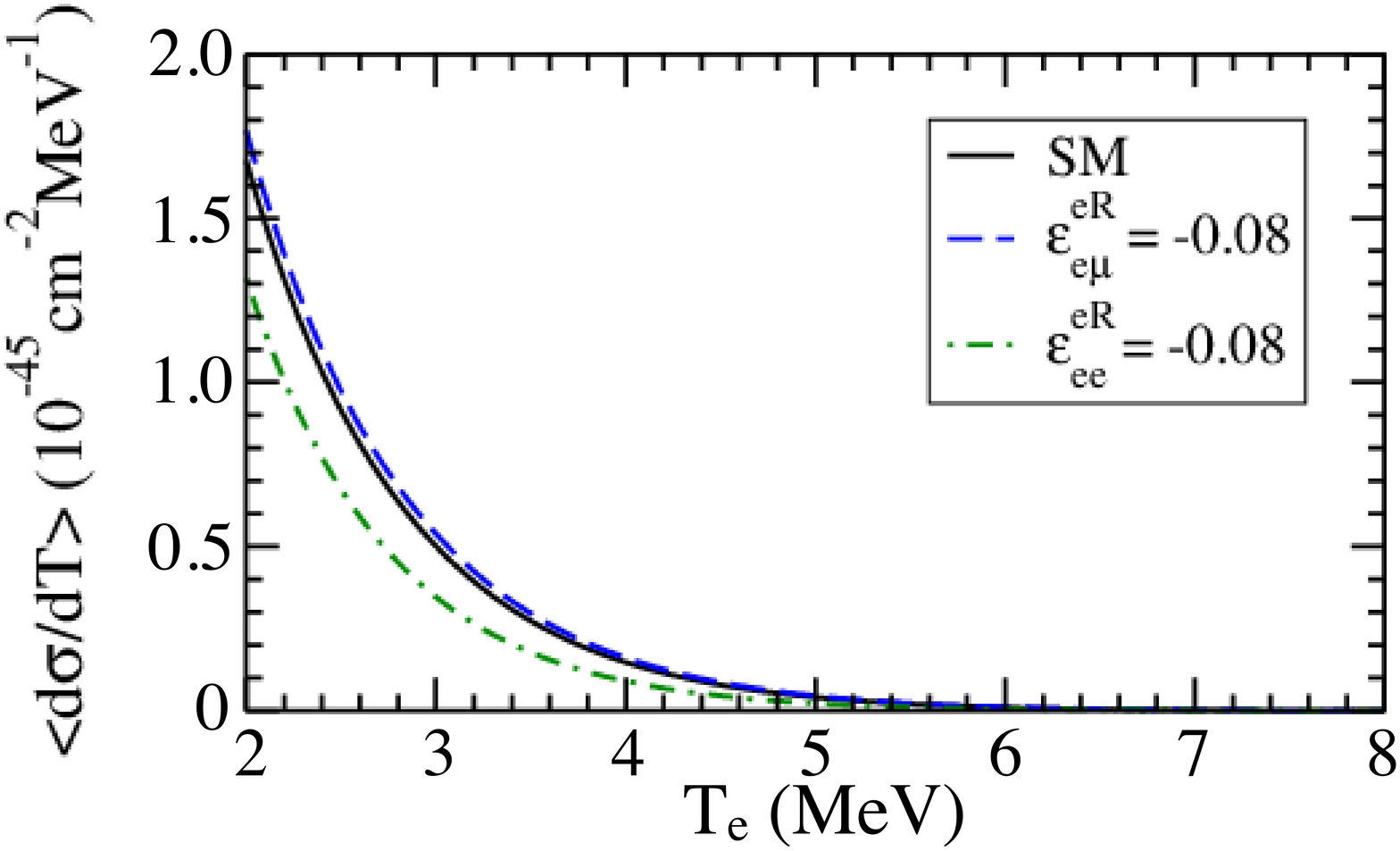}
\vglue -1.30cm
\caption{Averaged differential cross section for the electron
  anti-neutrino scattering off electrons for the SM case (black solid
  line), for a flavor changing NSI (blue dashed line), and for a
  flavor conserving NSI (green dashed dotted line).  The reactor
  anti-neutrino flux has been considered in order to integrate the
  anti-neutrino cross section over the appropriate neutrino energy
  range.}
\label{fig:reactor_xs}
\end{figure}

This is illustrated in Fig.~\ref{fig:reactor_xs}, where we show the
differential cross section for anti-neutrino electron scattering,
$\bar{\nu}_ee^-\to\bar{\nu}_ee^-$, averaged over a typical
anti-neutrino reactor spectrum~\cite{Mueller:2011nm,Huber:2004xh}. The
plot is given in terms of the electron recoil energy, $T_e$, for an
energy window relevant for an anti-neutrino detector such as
TEXONO~\cite{Deniz:2009mu,Wong:2006nx}. In the plot, the prediction
for the SM cross section is shown, as well as that for the NSI one.
For both flavor changing and flavor conserving NSI, the same negative
value of the parameters are used; this illustrates how flavor diagonal
NSIs have more impact on detection signals than flavor changing
parameters.

\begin{figure}[!h]
\centering
\includegraphics[width= 0.73\textwidth]{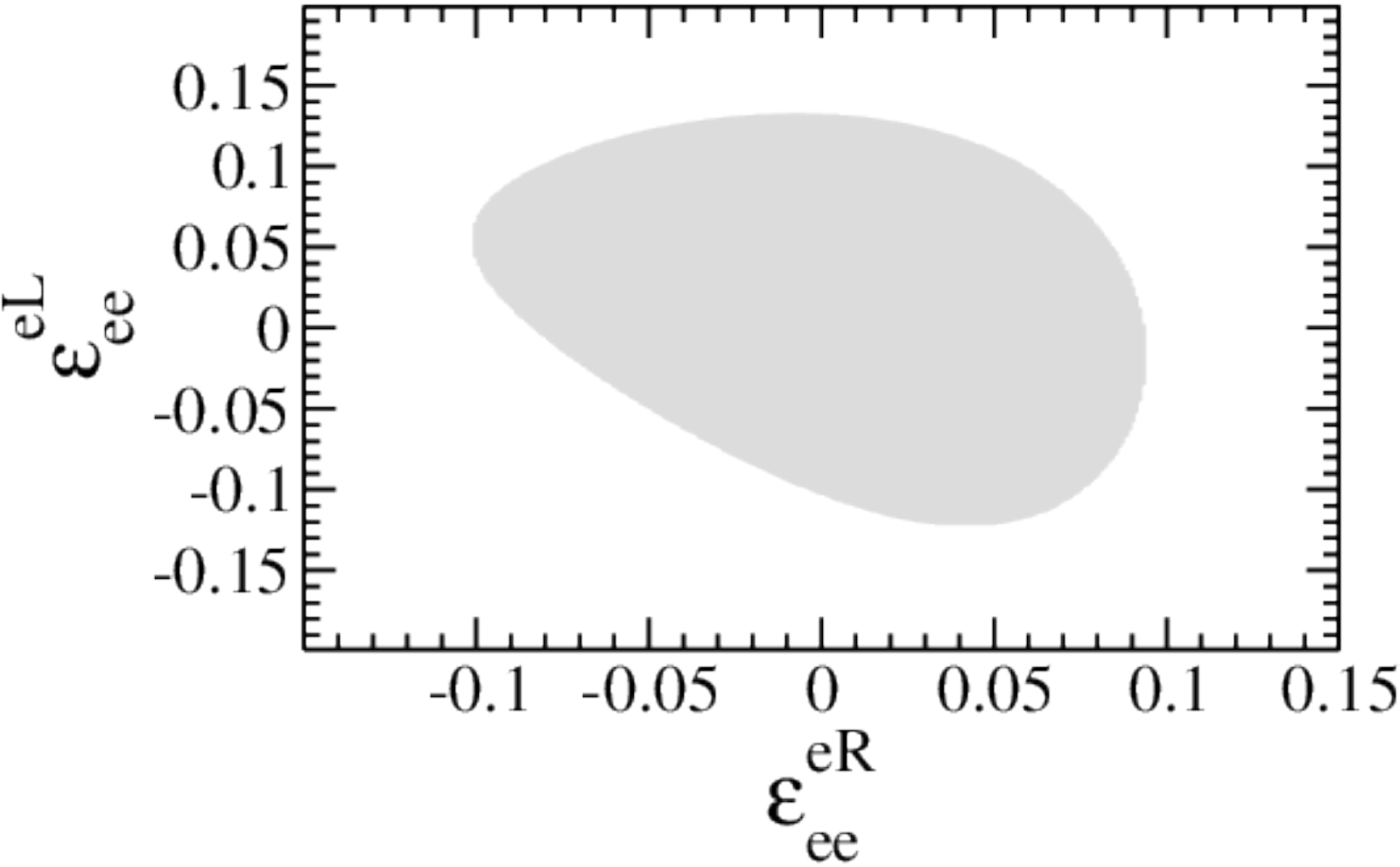}
\vglue -0.30cm
\caption{ Allowed region, at $90$ \% CL, for diagonal NSI parameters,
  $\varepsilon_{e e}^{L,R}$, from a combined analysis of TEXONO
  reactor anti-neutrino and LSND neutrino electron scattering off
  electrons.  }
\label{fig:chi-nue}
\vglue -0.50cm
\end{figure}

The NSI parameters for this reaction can be constrained by
considering, for example, the data from the TEXONO collaboration,
which use $\bar{\nu}_e e$ scattering as the detection signal. We have
updated the analysis reported by the TEXONO
collaboration~\cite{Deniz:2009mu}, including the new predicted
spectrum~\cite{Mueller:2011nm,Huber:2004xh} and radiative
corrections~\cite{Bahcall:1995mm} in order to obtain new constraints
for these parameters. We have also combined the results of this
analysis with the constraints coming from the $\nu_e e$ scattering
measurements reported by the LSND
collaboration~\cite{Auerbach:2001wg}.  By combining these two
experiments we can obtain stronger bounds both on left and right NSI
parameters, taking advantage of the different chirality of both
neutrino experiments.  The result of this new analysis is shown in
Fig.~\ref{fig:chi-nue} for the diagonal parameters
$\varepsilon_{ee}^{L,R}$, and is also shown in
Table~\ref{tab:constraint-e-det} along with other current
constraints.

Previous works involving the neutrino scattering off
electrons obtained constraints from a combined analysis of reactor
neutrino experiments. There are different experiments that involve an
electron (anti)neutrino flux, such as reactor neutrinos
(TEXONO~\cite{Deniz:2009mu}, MUNU~\cite{Daraktchieva:2003dr},
Rovno~\cite{Derbin:1993wy}, Krasnoyarsk~\cite{Vidyakin:1992nf},
Irvine~\cite{Reines:1976pv}) and accelerator neutrinos
(LSND~\cite{Auerbach:2001wg} and LAMPF~\cite{Allen:1992qe}). 

We show in Table \ref{tab:constraint-e-det} 
the summary of the constraints from these analysis.  
We prefer to show in this table results obtained by different
research groups since the analysis may have different assumptions that
can be important in the interpretation of the parameters. 
These results consider either one or two parameters at a time, while all
other parameters equal to zero; in some cases, especially for flavor
diagonal couplings, the correlation between these two parameters is
important.

We also show in Table \ref{tab:constraint-e-det} 
the constraints obtained from muon (anti)neutrino fluxes, 
based on the results coming 
from the CHARMII experiments~\cite{Vilain:1994qy}. 
Although there is no man made tau neutrino sources, 
it is possible to constrain
these interactions if one considers the LEP measurements of the
process~\cite{Berezhiani:2001rs} $e^+e^-\to\nu\bar\nu\gamma$ where tau
neutrinos appear as part of this inclusive reaction, or to consider
the solar neutrino flux that also includes a tau neutrino
component~\cite{Bolanos:2008km}; such constraints are also shown in
Table \ref{tab:constraint-e-det}.
%
\begin{table}[h!]
\vskip -.2cm
\centering
\caption{Constraints on the detection NSI couplings at
  90\% C L.  for the interaction of neutrinos with electrons}
\vskip .2cm
\label{tab:constraint-e-det}
\begin{tabular}{c|ll|ll}
\hline\hline
  & \multicolumn{2}{c|}{one parameter} &  \multicolumn{2}{c}{two parameter}   \\
\hline 
 $\varepsilon^{eL}_{ee}$ & $(-0.021,0.052)$~\cite{Bolanos:2008km} & 
                                                                & 
                         $( -0.02, 0.09)$~\cite{Forero:2011zz}  &
                         $( -0.036, 0.063)$~\cite{Bolanos:2008km}\\
 $\varepsilon^{eR}_{ee}$ & $(-0.07,0.08)$~\cite{Deniz:2010mp} & 
                         $( -0.08, 0.09)$~\cite{thiswork} & 
                         $( -0.11, 0.05)$~\cite{Forero:2011zz}  &
                         $( -0.10, 0.09)$~\cite{thiswork}  \\

\hline
 $\varepsilon^{eL}_{\mu\mu}$ & $(-0.03,0.03)$~\cite{Davidson:2003ha} & 
                         $( -0.03, 0.03)$~\cite{Barranco:2008}&
                                                                &
                         $( -0.033, 0.055)$~\cite{Barranco:2008}
                                                                 \\ 
 $\varepsilon^{eR}_{\mu\mu}$ & $(-0.03,0.03)$~\cite{Davidson:2003ha} & 
                         $( -0.03, 0.03)$~\cite{Barranco:2008} & 
                                                                &
                         $( -0.040, 0.053)$~\cite{Barranco:2008}  
                                                                 \\ 
\hline
 $\varepsilon^{eL}_{\tau\tau}$ & 
                         $(-0.16,0.11)$~\cite{Bolanos:2008km} & 
                         $( -0.46, 0.24)$~\cite{Barranco:2008} & 
                         $( -0.51, 0.34)$~\cite{Forero:2011zz}  & 
                         $( -0.16, 0.11)$~\cite{Bolanos:2008km} \\
 $\varepsilon^{eR}_{\tau\tau}$ &                                   & 
                         $( -0.25, 0.43)$~\cite{Barranco:2008} & 
                         $( -0.35, 0.50)$~\cite{Forero:2011zz}  &
                         $( -0.4, 0.6)$~\cite{Barranco:2008}\\
\hline \hline
 $\varepsilon^{eL}_{e\mu}$ &                                      & 
                           $(-0.13, 0.13)$~\cite{Barranco:2008} & 
                         $( -0.53, 0.53)$~\cite{Berezhiani:2001rs}  &
                          \\ 
 $\varepsilon^{eR}_{e\mu}$ & $(-0.19,0.19)$~\cite{Deniz:2010mp} & 
                         $(-0.13, 0.13)$~\cite{Barranco:2008} & 
                         $( -0.53, 0.53)$~\cite{Berezhiani:2001rs}  &
                          \\ 
\hline
 $\varepsilon^{eL}_{e\tau}$ & $(-0.4,0.4)$~\cite{Davidson:2003ha} & 
                         $( -0.33, 0.33)$~\cite{Barranco:2008}&
                         $( -0.53, 0.53)$~\cite{Berezhiani:2001rs}&
                          \\ 
 $\varepsilon^{eR}_{e\tau}$ & 
\multicolumn{2}{c|}{$( -0.28, -0.05)$\  \ and\ \ 
 ( 0.05, 0.28)~\cite{Barranco:2008} }
&                         $( -0.53, 0.53)$~\cite{Berezhiani:2001rs}  &
                          \\ 
                          & $(-0.19,0.19)$~\cite{Deniz:2010mp} & 
                          & 
                          &
                          \\ 
\hline
 $\varepsilon^{eL}_{\mu\tau}$ & $(-0.1,0.1)$~\cite{Davidson:2003ha} & 
                         $(-0.1, 0.1)$~\cite{Barranco:2008} & 
                         $( -0.53, 0.53)$~\cite{Berezhiani:2001rs}  &
                          \\ 
 $\varepsilon^{eR}_{\mu\tau}$ & $(-0.1,0.1)$~\cite{Davidson:2003ha} & 
                         $( -0.1, 0.1)$~\cite{Barranco:2008} & 
                         $( -0.53, 0.53)$~\cite{Berezhiani:2001rs}  &
                          \\ 
\hline
\end{tabular}
\end{table}

In order to get constraints on the NSI of neutrinos with $d$~type
quarks, it is necessary to study experiments such as CHARM,
CDHS~\cite{Allaby:1987vr,Blondel:1989ev} and, more recently, by
NuTeV~\cite{Zeller:2001hh}. They have measured the cross section for
the scattering of electron and muon neutrinos off quarks. For the case
of NuTeV, there have been a long discussion about a discrepancy of the
measured cross section with the SM prediction. After the revaluation
of the predicted sea contributions from the $c$ quark, it has been
possible to solve this puzzle~\cite{Ball:2009mk,Bentz:2009yy};
currently NSI suggested by the NuTeV experiment are considered as
consistent with zero.  

Table~\ref{tab:constraint-d-det} shows the summary of the 
constraints for the $d$ quark NSI coming from these
experiments. Constraints coming from charge lepton flavor conversion,
such as $\mu\to e\gamma$ or $\mu \to e$, have not been considered
here. They always involve, at some level, a one loop dressing of the
neutrino vertex; therefore, they will always be model dependent. The
readers interested in such constraints can see, for example, those
reported in~\cite{Papoulias:2013gha,Biggio:2009kv}.

\begin{table}[h!]
\centering
\caption{Constraints on the detection NSI couplings at 90\% C L. 
for the interaction of neutrinos with quarks}
\vskip .2cm
\label{tab:constraint-d-det}
\begin{tabular}{cr}
\hline\hline
NSI parameters & Bounds \ Ref. \\
\hline 
 $\varepsilon^{dL}_{ee}$ & $(-0.3,0.3)$~\cite{Davidson:2003ha} \\ 
 $\varepsilon^{dR}_{ee}$ & $(-0.6,  0.5)$~\cite{Davidson:2003ha} \\
\hline
 $\varepsilon^{dL}_{\mu\mu}$ &    $(-0.005, 0.005)$~\cite{Escrihuela:2011cf} \\
 $\varepsilon^{dR}_{\mu\mu}$ &   $(-0.042, 0.025)$~\cite{Escrihuela:2011cf} \\
\hline\hline
$\varepsilon^{dL}_{\mu e}$ 
& $(-0.023,0.023)$~\cite{Escrihuela:2011cf}  \\
$\varepsilon^{dR}_{\mu e}$
& $(-0.036,0.036)$~\cite{Escrihuela:2011cf}  \\ 
\hline
 $\varepsilon^{dL}_{e\tau}$ & $(-0.5, 0.5)$~\cite{Davidson:2003ha} \\
 $\varepsilon^{dR}_{e\tau}$ & $(-0.5, 0.5)$~\cite{Davidson:2003ha} \\
\hline
 $\varepsilon^{dL}_{\mu\tau}$ & $(-0.023,0.023)$~\cite{Escrihuela:2011cf}  \\
 $\varepsilon^{dR}_{\mu\tau}$ & $(-0.036,0.036)$~\cite{Escrihuela:2011cf}  \\ 
\hline 
\end{tabular}
\end{table}

\section{Future Prospects}
\label{sec:future}
There are several experimental proposals that plan to improve the current 
knowledge of neutrino properties. 
Therefore, there is plenty of room to 
improve the sensitivity to NSI in the near future. Some of these experimental 
set-ups are discussed below, showing the future perspectives for 
different types of experiments. Again, the discussion is divided into 
propagation and detection NSI. 

It is important to notice that, besides the need for more restrictive
bounds on NSI parameters, it is also necessary to solve the possible
confusion between standard and non-standard parameters. As has been
stated in the past~\cite{Huber:2002bi}, it is possible to have a
confusion in neutrino oscillation experiments between NSI parameters
and standard mixing angles, especially $\theta_{13}$; recently, this
subject has been discussed in the context of
solar~\cite{Palazzo:2011vg} and reactor
neutrinos~\cite{Girardi:2014kca}.  
The significant progress in improving the precision 
on $\theta_{13}$ will strongly restrict this possibility
in the near future. 

Another important topic in this direction is that of the robustness of
the solar neutrino data against NSI. It might be possible that large
NSI effect give rise to a dark-LMA solution without contradicting any
current experimental result~\cite{Miranda:2004nb}. This solution has
persisted as a plausible
picture~\cite{Escrihuela:2009up,Gonzalez-Garcia:2013usa}.  A recent
study on the future combined data of JUNO~\cite{Li:2013zyd}, 
RENO-50~\cite{Kim:2014rfa} and 
NOvA~\cite{Ayres:2004js} has discussed the perspectives to exclude
this solution~\cite{Bakhti:2014pva}.

\subsection{perspectives for NSI in propagation}

Different experimental set-ups (proposed to increase the precision for
the determination of the neutrino oscillations parameters) have been
considered in the last years.  The need for a better knowledge of the
Kobayashi-Maskawa type CP phase in the lepton sector as well as the
mass ordering is certainly a major motivation for these proposals.
Currently ongoing experiments such as T2K and NOvA may improve
somewhat the current bounds on some NSI parameters but probably not so
much, especially for that coming from the $\nu_e (\bar{\nu}_e)$
appearance mode due to relatively small statistics.

Hyper-Kamiokande~\cite{Abe:2011ts,Group:2014oxa} is an interesting
proposal that expects to improve the sensitivity to NSI, by using the
atmospheric neutrino
data~\cite{Fukasawa:2015jaa,Yasuda:2015lwa,Fukasawa:2014ota}; their
expectations for the normal hierarchy case are particularly appealing.
On the other hand, for the case of the LBNE~\cite{Adams:2013qkq} and
LBNO~\cite{Patzak:2012jp} proposals, the expected sensitivity to NSI
is also encouraging, especially for the flavor changing case of
$\epsilon_{\mu\tau}$ and $\epsilon_{e\mu}$~\cite{Huber:2010dx}.

Finally, important constraints are expected from the IceCube Deep Core
and PINGU experiments. These are extensions of the IceCube experiment
focused on a lower energy range. In this case, the expectations to
constrain the flavor changing NSI parameter $\epsilon_{\mu\tau}$ could
reach the one percent level. For the flavor diagonal case, it
might be possible to obtain information about the elusive parameter
$\epsilon_{\tau\tau}$~\cite{Esmaili:2013fva,Choubey:2014iia,Mocioiu:2014gua}.

\subsection{perspectives for NSI in detection}
The future neutrino oscillation experiments will also be sensitive to
the NSI parameters through a detection effect. For instance, for the
case of the proposed JUNO~\cite{Li:2013zyd} and 
RENO-50~\cite{Kim:2014rfa}  an improvement to the
constraints on $\epsilon_{e\mu}$ and $\epsilon_{e\tau}$ is
expected~\cite{Ohlsson:2013nna}.

For the case of the interaction of electron neutrinos with electrons,
both ISODAR~\cite{Conrad:2013sqa} and LENA~\cite{Wurm:2011zn} proposals 
could give complementary information if
both proposals are done in the future. In the case of
ISODAR, it is proposed to use an intense
anti-neutrino $^8$Li source, with an anti-neutrino energy ranging up to
$14$ MeV, in combination with the KamLAND liquid
scintillator~\cite{Conrad:2013sqa}. 
The LENA proposal plans to use a neutrino Chromium
source, providing a monochromatic neutrino flux of energy, $E_\nu=
0.747$ MeV, located at the top of a $100$~kTon liquid scintillator
cylindrical detector~\cite{Wurm:2011zn}. 
These are not the only proposals for neutrino
electron scattering, 
but they illustrate the future potential of this experiments. 

The use of either a neutrino or anti-neutrino source leads to a better
determination of the left or right-handed couplings,
respectively. Therefore, if both neutrino and anti-neutrino experiments
are done in the future, there could be a good room for the improvement
of the NSI parameters. 

\begin{figure}[h!]
\vglue -0.6cm
\hglue 0.6cm
\includegraphics[width= 0.90\textwidth]{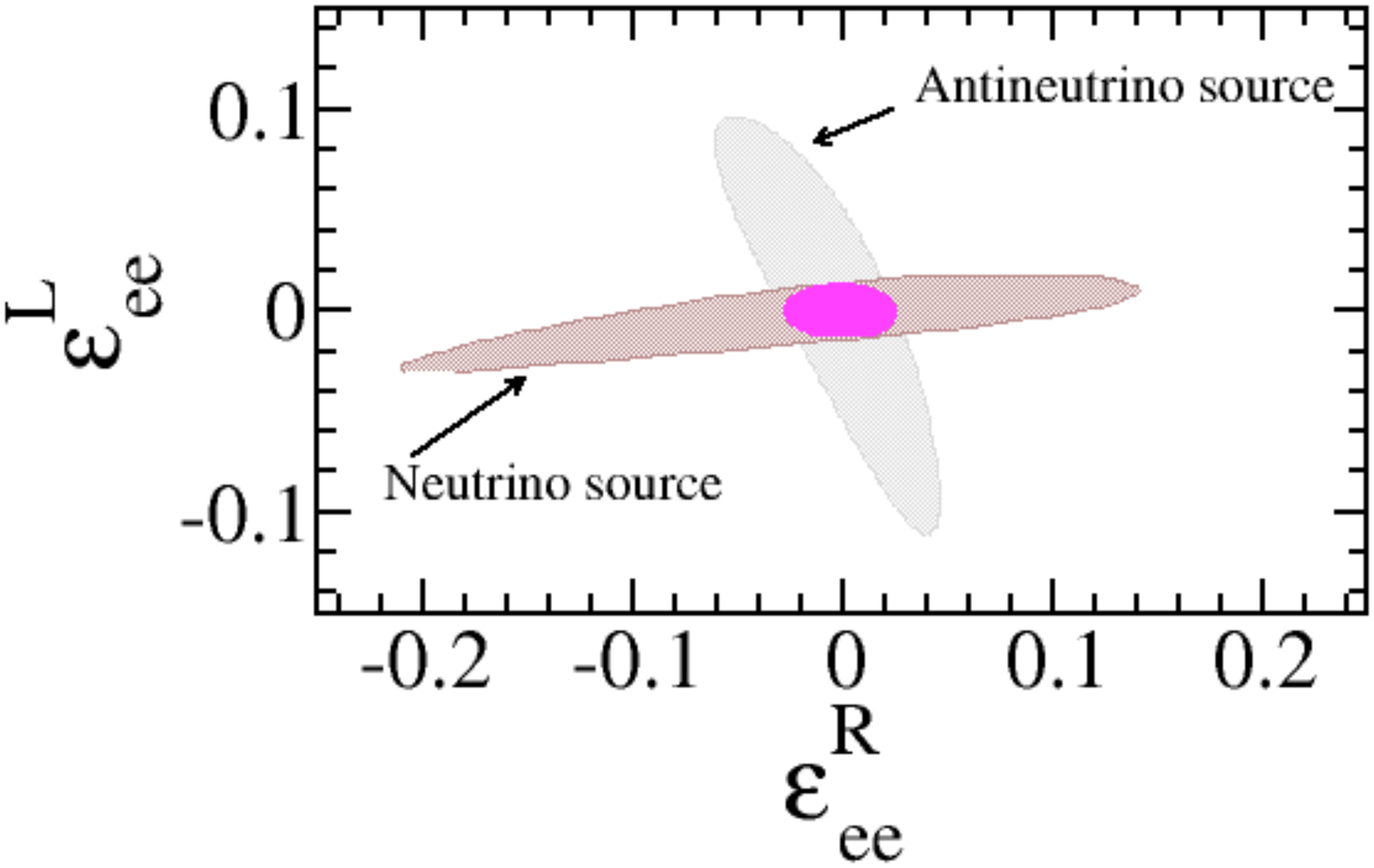}
\vglue -1.2cm
\caption{Expected sensitivity to NSI parameters
  $\varepsilon^{eL,R}_{ee}$ from ISODAR and LENA proposals. The result
  of a combined analysis is also shown.}
\label{nu-e:future}
\end{figure}

We illustrate this by showing, in
Fig.~\ref{nu-e:future}, the expected sensitivity for the case of a
neutrino artificial source in combination with the proposed  
LENA 
detector as has already been calculated in \cite{Garces:2011aa}, where
an expected total number of $1.9\times 10^{5}$ neutrino events and a
$5$~\% systematic error was considered.  The case of an anti-neutrino
source is also shown in Fig.~\ref{nu-e:future}. In this last case the
analysis developed in \cite{Conrad:2013sqa} has been closely followed.
The expected result from the combined analysis of both future
experiments is shown by the region filled by the magenta color.  We
show in the same figure one of the current constraints on NSI, coming
from the solar neutrino analysis~\cite{Bolanos:2008km}. It is possible
to see that there is room for improving these constraints by almost
one order of magnitude, especially if both experiments are realized.

Regarding the NSI of neutrinos with quarks, there are severals
proposals that could improve current bounds.  An interesting proposal
is that of neutrino coherent scattering off
nuclei~\cite{Freedman:1973yd}.  After seminal works on the
construction of these type of
detectors~\cite{Drukier:1983gj,Cabrera:1984rr}, there was a renewed
interest in the previous decade~\cite{Barbeau:2002fg,Wong:2005vg}. At
the same time, the sensitivity to NSI parameters and new physics
searches was also
noted~\cite{Barranco:2005yy,Barranco:2007tz}. Different experimental
set-ups have already been considered, either using a reactor
anti-neutrino flux~\cite{Wong:2005vg}, and spallation
source~\cite{Collar:2014lya,Akimov:2013yow,Scholberg:2009ha}, beta
beams~\cite{Espinoza:2012jr,Lazauskas:2007va}, or pion
decay~\cite{Brice:2013fwa,Anderson:2011bi,Scholberg:2005qs}. These
experiments could have an excellent sensitivity to the NSI. In
particular, the TEXONO proposal has been studied in the past, using
either a $^{76}$Ge or $^{28}$Si as a detector; an improvement of
even one order of magnitude could be achieved in this
case~\cite{Barranco:2005yy}.  Other possible nuclei have also been
studied~\cite{Papoulias:2013gha} such as $^{48}$Ti and $^{27}$Al.

\section{Conclusions}
\label{sec:conclusions}
Neutrino experiments have shown the existence of a new sector beyond
the Standard Model because of the experimental evidences of nonzero
neutrino masses that is already part of the current knowledge on
particle physics.  The mass and mixing of lepton sector turned out to
be non-trivial, very different from that of the quark sector.
While our knowledge on neutrino properties 
are continuously improving, 
the theoretical explanation of 
the neutrino mixing and the neutrino mass
pattern is still an open question.

In this context, possible presence of the non-standard interaction of
neutrinos and its impact was discussed in this brief review 
from a phenomenological point of view, describing the current 
status on the search for new physics coming from NSI. 

So far there is no experimental evidence or indication of the presence
of NSI and there exist only the constraints, which are summarized in
Tables \ref{tab:Table-NSI-propagation},~\ref{tab:constraint-e-det} and
\ref{tab:constraint-d-det} where we show the lower and upper bounds of
the NSI parameters.  Ongoing as well as proposed near future neutrino
experiments are expected to improve considerably the NSI bounds or may
indicate the presence of NSI.

\section*{Acknowledgments}
This work has been supported by CONACyT grant 166639 and EPLANET,
Conselho Nacional de Ci\^encia e Tecnologia (CNPq), 
and Funda\c{c}\~ao de Amparo \`a Pesquisa do Estado do 
Rio de Janeiro (FAPERJ).

\end{document}